\documentclass[10pt,twocolumn]{hdthep}
\usepackage{epsfig}
\usepackage{amsmath,amssymb}
\usepackage{graphicx}
\usepackage{subfigure}

\def\frac#1#2{\mathinner{#1\over#2}}

\begin{document}

\hdthep{01-32}
\author{Michael Doran\footnote{\tt doran@thphys.uni-heidelberg.de},\,\   Jan-Markus Schwindt\footnote{\tt schwindt@thphys.uni-heidelberg.de}  \,\ and  Christof Wetterich\footnote{\tt C.Wetterich@thphys.uni-heidelberg.de} }
\title{Structure Formation and the Time  Dependence of Quintessence }

\abst{We discuss the influence of dark energy on structure formation,
  especially the effects on $\sigma _8$. Our interest is particularly focused on
  quintessence models with time-dependent equation of state and
  non-negligible quintessence component in the early universe. We
  obtain an analytic expression for $\sigma _8$ valid for a
  large class of dark energy models. We conclude that structure
  formation is a  good indicator for the history of dark energy
  and use our results to set constraints on quintessence models.}

\maketitle

\section{Introduction}

\newcommand{\omd}{\Omega _{\rm d}}
\newcommand{\sa}{\sigma _8}
\newcommand{\omdsf}{\bar{\Omega} _{\rm d} ^{\rm sf}}
\newcommand{\omdn}{\Omega _{\rm d}^0}
\newcommand{\omdeq}{\sqrt{1-\bar{\Omega}_{\rm d} (\aeq)}}
\newcommand{\omdroot}{\sqrt{1-\bar{\Omega}_{\rm d} (a)}}
\newcommand{\aeq}{a_{\rm eq}}
\newcommand{\adec}{a_{\rm dec}}
\newcommand{\atr}{a_{\rm tr}}
\newcommand{\wda}{w_{\rm d}}
\newcommand{\wdan}{w_{\rm d}^0}
\newcommand{\kmax}{k_{\rm max}}
\newcommand{\keq}{k_{\rm eq}}

\subsection{Dark Energy}

There is evidence for dark energy contributing up to about 70\% of the
total energy of the universe \cite{Riess:1998cb,
  Perlmutter:1999np,Netterfield:2001yq}.  The nature of dark energy is
an open question, a cosmological constant or a dynamical scalar field
\cite{Wetterich:1988fm,Peebles:1988ek,Ratra:1988rm} called
quintessence \cite{Caldwell:1998ii} being two major options.  The
interest in quintessence arises from the possibility that the enormous
fine-tuning problems plaguing a cosmological constant can partially be
cured by some quintessence models. However, telling the difference
between a cosmological constant and quintessence or between different
quintessence models is complicated because of the non-genericness of
quintessence.  

If quintessence couples only gravitationally to matter,
the only way of detection is possibly
the exploration of its time dependent energy density and equation of
state, as the relative energy density fluctuations 
$\delta \rho_{\rm d} / \rho_{\rm d}$ within the horizon are negligible
\cite{Ferreira:1998hj}.  In order to find this time dependence,
measurements of different epochs are necessary.
For that reason, the interplay between quintessence and 
nucleosynthesis \cite{Bean:2001wt}, cosmic microwave background (CMB)
\cite{Amendola:2001ub,Doran:2001ty,Corasaniti:2001mf}, weak lensing
\cite{Huterer:2001yu} and Supernovae Ia data
\cite{Huterer:2000mj,Weller:2001gf} have recently been explored.

Also, the theory of structure formation can in principle test the
history of quintessence in the large range of redshift $z \in [0,\ 
10^4]$.  As has been noticed in \cite{Ferreira:1998hj,
  Skordis:2000dz}, the presence of dark energy can influence the
growth of structure in the universe from matter radiation equality
onwards.  In particular, $\sa$, the rms density fluctuations
fluctuations averaged over $8 h^{-1} \textrm{Mpc}$ spheres, is a
sensitive parameter.  Until now, a quantitative understanding of the
effect of quintessence on $\sa$ has been missing.  This paper aims to
fill this gap.

Supernovae Ia observations
\cite{Riess:1998cb, Perlmutter:1999np} indicate that the equation of
state $\wda \equiv p_d / \rho _d$ of dark energy is negative today. 
This 
gives rise to the aforementioned fine-tuning problem: We have
\begin{equation}
 \frac{\omd(a)}{\Omega _m (a)}\propto a^{-3 \tilde{w}_d},
\end{equation}
where $a$ is the scale factor and $\tilde{w}_{\rm d}$ is an appropriate mean value for the equation 
of state. If $\wda$ has always been negative, like in the case of the 
cosmological constant, $\omd(a)$ has been extremely small in the early universe,
and its importance just today lacks a natural explanation. Scalar models with this property
can be constructed for an appropriate effective scalar potential \cite{Ratra:1988rm}. They often
involve, however, an unnatural tuning of parameters \cite{Hebecker:2001zb}. 
The problem can be surrounded if we assume that $\wda$ became negative relatively 
recently and $\rho _d$ has scaled in the past like radiation or matter.
We will call such models `models with early quintessence' and pay particular
attention to them.

The COBE \cite{Bennett:1996ce} normalization \cite{Bunn:1997da} 
of the CMB power spectrum determines $\sa$ for 
any given model by essentially fixing the fluctuations at decoupling. 
This prediction is to be compared to values of $\sa$ infered from other experiments, such as 
cluster abundance constraints which yield \cite{Wang:1998gt}
\begin{equation}\label{cluster}
 \sigma _8 =(0.5 \pm 0.1)\Omega _m^{-\gamma},
\end{equation} 
where $\gamma$ is slightly model dependent and usually  $\gamma\approx 0.5$.
A model where these two $\sa$ values do not agree can be ruled out.
Standard Cold Dark Matter (SCDM)  without dark energy
\footnote{and $h=0.65,\ n=1, \Omega_{\rm b}h^2 =0.021,\ \Omega_{\rm m}^0=1$}
 for instance gives  $\sa^{\rm cmb} \approx 1.5,\ \sa^{\rm clus.} \approx 0.5 \pm 0.1$  
and is hence incapable of meeting both constraints.

CMB measurements \cite{Netterfield:2001yq,Lee:2001yp} suggest that the universe 
is flat, \mbox{$\Omega\equiv\Omega _{\rm m} + \Omega _{\rm r} + \omd =1$} and we assume this throughout the paper.\footnotemark

\footnotetext[1]{We use here conventions where $0$ always denotes today's value of a quantity and
the subscript {\it m,\ r {\rm and} d}  denote matter, radiation and dark energy respectively.}

\subsection{Quintessence vs Cosmological Constant}
Our main result is an estimate of the CMB-normalized $\sa$-value for 
a very general class of 
Quintessence models Q just from the knowledge of  their ``background solution'' 
$[\omd(a),\ \wda(a)]$ 
and the $\sa$-value of the $\Lambda$CDM model $\Lambda$ with 
the same amount of dark energy today
 $\Omega _{\Lambda}^0=\omdn(\Lambda)$:
\begin{equation} \label{main}
 \frac{\sigma _8 (Q)}{\sigma _8 (\Lambda)}\approx
\left( a _{\rm eq}\right)^{ 3\, \omdsf / 5}
 \left(1-\Omega _{\Lambda}^0 \right)^{-\left (1+ \bar w ^{-1}\right)/5}
 \sqrt{\frac{\tau _0 (Q)}{\tau _0 (\Lambda)}}.
\end{equation} 
If Q is a model with `early quintessence', $\omdsf$ is an average value 
for the fraction of dark energy 
during the matter dominated era, before $\omd$ starts growing rapidly at 
scale \mbox{factor $a_{\rm tr}$:}
\begin{equation}
 \omdsf \equiv [ \ln{a_{\rm tr}}-\ln{\aeq} ]^{-1} \int_{\ln \aeq}^{\ln a_{\rm tr}} \omd(a)\  {\rm d} \ln a  .
\end{equation}
If Q is a model without early quintessence, $\omdsf$ is zero. The effective equation of state
 $\bar{w}$ is an
average value for $\wda$ during the time in which $\omd$ is growing rapidly:
\begin{equation}
\frac{1}{\bar{w}}=\frac{\int _{\ln a_{\rm tr}}^0 \omd(a)/w(a)\:  d \ln{a}}
                   {\int _{\ln a_{\rm tr}}^0 \omd(a)\:  d  \ln{a}}.
\end{equation}
In many cases, $\wdan$ can be used as an approximation to $\bar{w}$ since the
integrals are dominated by periods with large $\omd$. The scale factor at
matter radiation equality is
\begin{equation}\label{equality}
  \aeq=\frac{\Omega _{\rm r}^0}{\Omega _{\rm m}^0}= \frac{4.31 \times 10^{-5}}
  {h^2(1-\omdn)}.
\end{equation}
Finally, 
$\tau _0$ is the conformal age of the universe.
\mbox{Equation \eqref{main}}  in combination with
\eqref{cluster} can be used to make general statements about the consistency of quintessence
models with $\sa$-constraints.

\subsection{Structure Formation}
In linear approximation, the theory of structure formation describes the evolution of the 
energy density contrast $\delta$
\begin{equation}
  \delta(\mathbf{x},a)\equiv\frac{\delta\rho _m(\mathbf{x},a)}{\bar \rho _m(a)}
  =\frac{\rho _m(\mathbf{x},a)-\bar \rho _m(a)}{\bar\rho _m (a)},
\end{equation}
and its fourier transform
\begin{equation}
  \delta _k (a)\equiv V^{-1}\int _V \delta(\mathbf{x},a)\exp{(-i \mathbf{k}\cdot\mathbf{x})}d^3 x.
\end{equation}
Here, V is the integration volume and $\mathbf{k}$ and $\mathbf{x}$ are the comoving wave vector
and the comoving coordinate. The structure growth exponent $f$ is defined as
\begin{equation} \label{growth}
  f(a) \equiv \frac{d \ln{\delta _{k}(a)}}{d \ln{a}}, 
\end{equation} 
and is roughly $k$-independent for a wide range of $k$. One can use linearized General Relativity
in the synchronous gauge to compute $f(a)$. For sub-horizon modes in SCDM models, one obtains 
$f \rightarrow 0$ in the radiation and $f=1$ in the \mbox{matter eras.  \footnotemark}
We define the \emph{growth factor} $g$ of density perturbations between 
arbitrary $a_{1}\!<a_{2}$ as the ratio $\delta_k(a_{2}) / \delta_k(a_{1})$. With a suitably
defined average growth exponent $\bar f$, this is 
\begin{equation}\label{growthFactor}
g(a_1,a_2) \equiv \frac{\delta_k(a_{2}) }{\delta_k(a_{1})} = \left(\frac{a_{2}}{a_{1}}\right)^{\bar f}
\end{equation}
The density contrast $\sa$ is defined by
\begin{equation}
  \sa^2 \equiv\int _0 ^\infty \frac{dk}{2 \pi^2} k^2 \delta _k^2 (k)
  \left ( \frac{3j_1 (kr)}{kr} \right ) ^2
\end{equation}
with $r = 8h^{-1} {\rm Mpc}$.
\footnotetext{super-horizon modes grow as  $f=2,\ (f = 1)$ in the radiation (matter) era.}

\begin{figure}
\begin{center}
\includegraphics[scale=0.4, angle =-90]{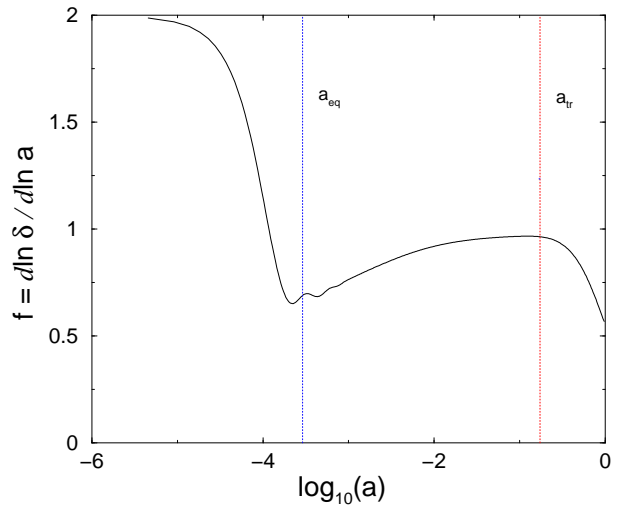} 
\caption{Time evolution of the growth exponent $f$ for mode $8h^{-1}\times k =1.95$,
shown as a function of $\rm log_{10}(a)$. One observes horizon crossing around $\rm log_{10}(a) = -4$
and $f$ near unity in the matter dominated era with a drop due to the increase of
dark energy in the present period. We indicate $\aeq$ and $\atr$ as specified in the 
text as vertical lines. The curve is obtained for leaping kinetic term quintessence with 
$h=0.65,\ \Omega_{\rm b}h^2=0.02,\ \omd^0=0.65,\ \omdsf=0.045$. 
}
\label{fPic}
\end{center}
\end{figure}

The modes with the highest weight from the $\sa$ window function entered
the horizon during the late radiation era. After horizon crossing
$f$ decreases and starts to grow again around
matter-radiation-equality (see Figure \ref{fPic}).

\subsection{Evolution Equations}
For simplicity, we set the reduced
Planck mass $M_{\bar P}=(8 \pi G)^{-1/2}$ to unity.
The Friedmann equation for a flat universe is then
\begin{equation} \label{friedman}
3H^2=\rho _m + \rho _r + \rho _d.
\end{equation} 
From this and the appropriate scaling of the different 
energy densities due to the expansion of the Universe, we get
\begin{equation}
  \frac{d \ln{H}}{d \ln{a}}=-\frac{1}{2}\left [3+3\wda(a) \Omega _d(a) +\Omega _r(a)\right]
   \equiv -\frac{1}{2}\tilde{n}(a),
\end{equation} 
where $\tilde{n}(a)=3,4$ for matter (radiation) domination.
The sub-horizon growth of density perturbations is governed by
\begin{equation} \label{df}
  \frac{df}{d \ln{a}}+f^2 +\left ( 2-\frac{1}{2}\tilde{n} \right ) f
  -\frac{3}{2}\Omega _m =0.
\end{equation}
If $\tilde{n}$ and $\Omega _m$ are constant (as e.g. in the Exponential Potential Model,
see section \ref{epm}, or in SCDM), $f$  approaches the solution
\begin{equation}
 f=\frac{\tilde{n}}{4}-1+\frac{1}{2}\sqrt{\left ( 2-\frac{\tilde{n}}{2} 
 \right ) ^2 +6 \Omega _m}.
\end{equation}
Time and conformal time are related to the scale factor by
\begin{equation}
  \frac{dt}{da}=\frac{d \tau}{d \ln{a}}=\frac{1}{aH}.
\end{equation}
Later on we will need an expression for $\tau(a=1) \equiv \tau_0$. Using the Friedmann equation we obtain
\begin{eqnarray} \label{age1}
 \tau(a)&=&\int^{a}_{0}\frac{da'}{{a'}^2 H(a')} \nonumber \\
 &=&2 \sqrt{3}\omdroot \frac{\sqrt{\rho _{\rm m}^0a+\rho _{\rm r}^0}-\sqrt{\rho _{\rm r}^0}}{\rho _{\rm m}^0}.
\end{eqnarray}
where 
\begin{equation}\label{age2}
 \omdroot \equiv \int^{a}_{0}\frac{da'\sqrt{1-\Omega _{d}(a')}}
   {\sqrt{\rho _{\rm m}^0a'+\rho _{\rm r}^0}}/\int^{a}_{0}\frac{da'}
   {\sqrt{\rho _{\rm m}^0a'+\rho _{\rm r}^0}}.
\end{equation} 

\subsection{CMB Normalization}
The CMB temperature anisotropies are related to the density perturbations at 
the time of decoupling. The dominant contribution on large scales is the  
Sachs-Wolfe \cite{Sachs:1967er} effect:
\begin{equation}
 \frac{\delta T}{T}(\adec)=\frac{\delta\rho}{3 \rho}(\adec).
\end{equation}
This (or indeed the refined method of \cite{Bunn:1997da}) is
used to fix $\delta _k (\adec)$ on scales $k<0.01 \: h \, {\rm Mpc}
^{-1}$.  The CMB is emitted from the surface of last scattering (SLS)
which has the comoving distance $d _{SLS}=\tau _0 - \tau _{dec}$ from
the observer. The knowledge of this distance is necessary for the
normalizing procedure because the measured angular CMB power spectrum
has to be converted to a momentum power spectrum, and so the
normalization depends on $\tau _0$ (see 
section \ref{influence}).

\section{The Influence of Dark Energy \mbox{on $\sigma _8$}}
\label{influence}
In this section, we compare structure formation in universes with dark
energy to that in SCDM and find five differences in the computation of
the CMB-normalized $\sa$-value. The first four effects concern the
growth of structure according to Equation \eqref{df}, whereas the fifth
affects the normalization of the matter power spectrum.\\[-1.2ex]

{\it Equality shift}: 
We have $\aeq\propto (1-\omdn)^{-1}$. If \mbox{$\omdn\approx 0.6$}, then
$\aeq$ is larger than in SCDM by a factor of $2.5$. Therefore $f$ starts growing much later for
the $\sa$-relevant modes, leading to a substantially lower $\sa$-value. This effect is the
strongest for many dark energy models. It would be difficult to compute it analytically,
because around equality too many physical processes play a role at the relevant scale
( e.g.\ horizon entering, decoupling, damped oscillation of radiation fluctuations). 
We circumvent the difficulty of computing this effect analytically by comparing models with the
\emph{same} dark energy content today and therefore identical values of $\aeq$.
It is then sufficient to determine $\sa$ numerically (by {\sc cmbfast}  \cite{Seljak:1996is}) for one model of this
class, e.g. $\Lambda$CDM.\\[-1.2ex]
  
{\it Matter depletion}: 
From Equation \eqref{df} we see that a decrease of $\Omega _m$ leads to a decrease of $f$.
We will discuss this effect analytically in the next section.\\[-1.2ex]

{\it Accelerated expansion}: 
Also from  Equation \eqref{df} we see that an accelerated expansion, i.e.\ a smaller value of 
$\tilde{n}$, leads to a decrease of $f$. The accelerated expansion typically affects only
a recent epoch in the evolution of the universe, and hence the effect will be rather small.\\[-1.2ex]

{\it Shift in horizon crossing}:
Due to the different expansion
history, a mode $k$ enters the horizon at a different scale factor
than in SCDM. As the equality shift, this is difficult to calculate
analytically. Once again, we partially evade this difficulty by
comparing models with the same $\omdn$. Numerically, we find the
residual effect to be small compared to the other effects and hence we
will neglect it.\\[-1.2ex]

{\it Normalization shift}: 
A universe with dark energy is typically about 30 to 60 \% older than
a  SCDM universe. This means that the distance $\tau _0 -\tau _{dec}$ to the SLS
is larger than in SCDM. Thus, the measured angular
temperature correlations correspond to momentum space correlations with smaller 
\mbox{ $k$:  $(\delta T)_k=(\delta T)_{k'}(SCDM)$},
\begin{equation}
 \frac{k'}{k}=\frac{\tau _0 -\tau _{\rm dec}}{\tau _0(SCDM)-\tau _{\rm dec}(SCDM)}
 \approx\frac{\tau _0}{\tau _0(SCDM)}.
\end{equation}
From the Sachs-Wolfe effect,
the CMB temperature perturbations are proportional to the density perturbations which on super-horizon
scales are determined by the initial power spectrum 
\begin{equation}
P(k) \equiv \delta _k^2=A_{\rm q} \: k^n = A_{\rm q} k^{\prime\,  n}  \times (k/k')^n 
= A_{\textsc{scdm}} k^{\prime\,  n}
\end{equation}
with spectral index $n$. Hence, we get for the ratio of perturbations
\begin{equation}\label{norm}
 \frac{\delta _k (\adec)}{\delta _k (\adec,SCDM)} = \left( \frac{k'}{k} \right) ^{n/2} 
 \approx\left ( \frac{\tau _0}{\tau _0(SCDM)}\right ) ^{n/2}.
\end{equation} 

With $n \approx 1$ this accounts for the last factor in Equation \eqref{main}.

\section{Effective Models}
In this section we show how generic dark energy models can effectively
be described as an appropriate combination of quintessence with an
exponential potential (for small $a$) and a dark energy model with
constant equation of state $\wda$ (for large $a$). We start by investigating
the two `pure cases' separately. 
\subsection{The Exponential Potential} \label{epm}
Quintessence with an Exponential Potential (EP) \mbox{$V(\phi)=e^{-\lambda\phi}$}
(and standard kinetic term)
has been investigated in 
\cite{Wetterich:1988fm, Ratra:1988rm,Wetterich:1995bg,Ferreira:1998hj}.
For $\lambda > 2$,  the quintessence field is forced into an
attractor solution with $\omd =4/\lambda^2$ during radiation domination and 
$\omd=3/\lambda^2$ during matter domination. So $\omd$ is constant during structure 
formation. The expansion history of the universe (and hence $\tau$) is almost unchanged
compared to SCDM, and 
during matter domination we have $\wda=w_m=0$.
Present observations on $\omdn$ and $\wdan$ suggest that this model needs to be modified
for large $a$, at least around the present epoch $a \approx 1$.
\emph{Matter depletion} is the dominant effect on structure formation.
The structure growth exponent (see \cite{Ferreira:1998hj})
\begin{equation}\label{ferf}
 f=\frac{-1 + \sqrt{25-24 \Omega _d}}{4}=1-\frac{3}{5}\omd + O(\omd^2),
\end{equation}  
is smaller than in SCDM and reduces $\sa$ correspondingly. For small 
$\omd$ this amounts to a change in  $\sa$ by a factor of 
\begin{equation} \label{epfakt}
 \frac{\sa (EP)}{\sa (SCDM)}\approx \left ( \frac{1}{\aeq} \right ) ^{-3\omd/5}.
\end{equation}
The \emph{Equality shift} lowers $\sa$ even more. 
For the EP model, it  follows that 
10\% Quintessence lowers $\sa$ by about 50\%.
If a Quintessence model is different from the EP, but $\omd$ is relatively small all the 
time and does not vary too fast, its effect on structure formation will be almost like
an EP model. According to Equations \eqref{growth}, \eqref{ferf}
we have to replace $f$ and therefore $\omd$ in this case by its 
logarithmic mean value 
\begin{equation}
 \omdsf  \equiv - \left (\ln \aeq \right)^{-1} \int _{\ln \aeq}^0 \omd(a) d \ln a ,
\end{equation}
for structure formation. Thus, Equation \eqref{epfakt} 
remains valid if we substitute $\omdsf$ for $\omd$.

\subsection{Dark Energy with Constant Negative Equation of State}
Dark energy models with constant equation of state (CES) $\wda<0$ have
been investigated e.g. in \cite{Caldwell:1998ii, Wang:1998gt}. We wish
to extend the analytical discussion by making some simplifications.
The CES models have the property that the dark energy becomes
important just in the present epoch. The matter depletion is not as
important as in the EP case, because the decrease of $f$ just started
recently. If $w$ is closer to zero, the matter depletion becomes
stronger, because the dark energy component became important earlier
in the past.  Current data favors models with  $\omdn \in [0.5,\ 0.7]$, so the
equality shift is very strong. The expansion history of the universe
is changed giving rise to the \emph{accelerated expansion} and
\emph{normalization shift} effects.

{\vspace{1ex} \it Matter depletion and accelerated expansion:}
Numerically we find that the approximation
$f(a)\approx 1-\frac{3}{5}\omd(a)$
is still valid to about 5\% in CES models, even when $\omd$ is as large as $0.6$. 
As $\omd(a) / \Omega _m(a)\propto a^{-3w}$, there is no dark energy 
contribution at early times when the radiation component is significant. Conversely,
radiation is negligible when dark energy contributes and hence
\begin{equation}\label{darkA}
 \omd(a)=\frac{\omdn}{(1-\omdn)a^{3w}+\omdn}.
\end{equation}
Now, to quantify the difference in structure growth compared to SCDM we  
fix an $a_{\rm tr}$ which lies 
in the matter era in both the SCDM and the CES model and at  which dark energy 
contribution is small. The appropriate averaged growth exponent from $a_{\rm tr}$ 
to today is then given by
\begin{eqnarray}
\bar f &=& -\left [ \ln(a_{\rm tr})\right]^{-1}  \times \int _{\ln a_{\rm tr}}^{0} \left[ 1 - \frac{3}{5}\omd(a) \right] \ d \ln a \nonumber \\
&\approx&  1 +  \frac{3}{5} \int _0^1  \omd \ \frac{{\rm d}a}{a}  / \ln(a_{\rm tr})\nonumber\\ 
& =& 1 +   \frac{1}{5w}\ln{(1-\omdn)} / \ln(a_{\rm tr}) .
\end{eqnarray}
Using  Equation \eqref{growth} we find that  the change in the growth
factor \eqref{growthFactor} 
\begin{equation}\label{gRatio}
  \frac{g(\atr,a=1; \textsc{ces})}{g(\atr,a=1; \textsc{scdm})} 
  = { a^{ (1- \bar f) }_{\rm tr} } =(1-\omdn)^{-1/(5w)}
\end{equation}  
is independent of $a_{\rm tr}$.

{{\vspace{1ex} \it Normalization shift:}}
According to Equation \eqref{norm},
we must compute the conformal age of the universe $\tau _0 =\tau(a=1)$. Neglecting the radiation
density, we obtain from Equations \eqref{age1} and \eqref{darkA}
\begin{equation}
 \tau _0 = \frac{2 \sqrt{3}}{\sqrt{\rho _{\rm m}^0}}
                     \:  F \left ( \frac{1}{2},\frac{-1}{6w},1-\frac{1}{6w},
                         \frac{-\omdn}{1-\omdn} \right ),
\end{equation}
where $F$ is the hypergeometric function $_2F_1$.

We  would now like to compare two CES models A and B which 
have the same $\omdn$ but different $w$.
Because $\omdn$ is the same, the equality shift 
is the same in both cases and cancels out. From
the other effects we get (cf. Equations \eqref{norm}, \eqref{gRatio})
\begin{equation}
 \frac{\sa(A)}{\sa(B)}\approx(1-\omdn)^{(w_B^{-1}- w_A^{-1})/5}
 \sqrt{\frac{F(w_A)}{F(w_B)}}. 
\end{equation}
For realistic values of $\omdn$ and $w$, this approximation is precise to about 5\%.

We next consider models where $\wda$ is time-dependent but always negative.
If $w$ does not vary rapidly, the difference does not become relevant as
long as $\omd$ is substantially larger than zero, and we can take today's value $\wda(a=1)$
as an approximation and consider the model as an CES model. If $w$ varies rapidly, we can
instead use
an average value of $\wda$ defined via
\begin{equation}
\frac{1}{\bar{w}}=\frac{\int _{\ln \aeq }^0 \Omega _d (a)/w(a)\: d\ln{a}}
                   {\int _{\ln \aeq}^0 \Omega _d (a)\: d \ln{a}}.
\end{equation}
         
\subsection{General Models}
We will now consider models with negative $w$
today but non-negligible quintessence in the early universe, i.e.
those with $\wda(a) \geq 0$ for small $a$. We call them models with
early quintessence (EQ).  Such models are particularly interesting
because they combine the naturalness properties of EP models with the
realistic late cosmology of CES models.  The difference between EQ and
the CES-like models is relevant for structure formation only in the
case that $\wda(a) \geq 0$ in an $a$-interval after equality (unlike
in k-essence \cite{Armendariz-Picon:2001ah} where this is only the
case in the radiation era) where   $\omd(a)$ is non-negligible. 
For the phenomenology of structure formation, we will  describe 
these models as a combination of EP-
and CES-models.  For any early quintessence model EQ, we pick a
certain scale factor $a_{\rm tr}$ at which the dark energy's equation
of state falls below $-0.25$ (although the precise value is not essential 
to our results).
For the effects related to the growth factor (matter depletion
and accelerated expansion), we consider the periods before and after
$a_{\rm tr}$ separately and multiply the growth factors arising from
both epochs. The time history of the growth exponent $f(a)$ for 
a typical model in this class is shown in figure \ref{fPic}.

 For $a<\atr$, we consider this EQ model as an effective exponential
model EP with
\begin{equation}
 \omdsf ({EP}) = \frac{\int _{\ln \aeq}^{\ln a_{\rm tr}} \Omega _d(a) d \ln a}{\ln{a_{\rm tr}}-\ln{\aeq}},
\end{equation}
while for $a>a_{\rm tr}$, we treat it as an effective CES model
\begin{equation}
\frac{1}{\bar{w}({CES})}=\frac{\int _{\ln a_{\rm tr}}^0 \Omega _d (a)/w(a)\: d\ln{a}}
                   {\int _{\ln a_{\rm tr}}^0 \Omega _d (a)\: d \ln{a}}.
\end{equation}
The effective total growth factor is obtained by multiplying the
expressions \eqref{epfakt} and \eqref{gRatio} with appropriate
modification of \eqref{norm}, replacing $1/\aeq$ by $a_{\rm tr}/\aeq$
and $\omd$ by $\omdsf$.  In general, $a_{\rm tr}$ will be relatively
close to unity if $\omdsf$ is non-negligible.

We are now in the position to derive Equation \eqref{main} by combining the results for
EP and CES models of the previous sections. For two general models A and B 
with the same $\omdn$ (and hence $\aeq$) but different histories i.e. different $w$ and $\omdsf$, we find
\begin{multline}\label{alles}
 \frac{\sigma _8 (A)}{\sigma _8 (B)}\approx
 \left ( \frac{a_{\rm tr}}{\aeq} \right ) ^{-3\left(\omdsf(A)-\omdsf(B)\right)/5}
 (1-\omdn)^{\left(\bar w_B^{-1}-\bar w_A^{-1}\right)/5} \\
 \times \sqrt{\frac{\tau _0 (A)}{\tau _0 (B)}}.
\end{multline} 
Inserting the values relevant for the cosmological constant model,
$\omdsf =0,\ \bar w(\Lambda)=-1$ and replacing $\atr$ with $a_0=1$ in
\eqref{alles} (thus neglecting the cutoff of the EP part), we obtain
Equation \eqref{main}.

The usefulness of this Equation (which is precise to about 5\%) lies
in the fact that we need to numerically compute the $\sa$-value of
only one model - e.g. $\Lambda$CDM - for a given $\omdn$.  From this,
$\sa$ of all other models with the same $\omdn$ can be estimated from
their background solution only.  We see that $\sa$ depends on the
three quantities $\omdn$, $\omdsf$ and $\bar w_{\rm d}$.  Here, the
dark energy today, $\omdn$, mainly contributes through the equality
shift and a small amount through the normalization shift.  The amount
of dark energy during structure formation, $\omdsf$ enters through
matter depletion in the EP-part of the model, and the late-time
equation of state $\bar w$ through matter depletion and normalization
shift in the CES-part of the model.  Hence $\sa$ is a very promising
quantity for constraining these quantities.

\section{Dependence of $\sa$ on other Parameters}
The density contrast $\sa$ can be a sensitive indicator of the
detailed properties of dark energy once the other cosmological
parameters are known.  For the present, our ability to constrain dark
energy models using $\sa$ is limited by the imperfect knowledge of
these parameters. We observe a certain degeneracy arising in
particular from $h$ and the spectral index $n$.

The Hubble parameter $h$ appears in the denominator of Equation
\eqref{equality} and hence strongly affects the equality shift. A
higher value of $h$ leads to a smaller $\aeq$ and so to a higher value
of $\sa$. On the other hand a higher value of $h$ gives a smaller
$\tau _0$, hence a smaller CMB normalization. The first effect is much
stronger than the second one.

The spectral index $n$ appears in the CMB normalization procedure, when the
measured large scale anisotropies 
($k \approx  10^{-2} h \: {\rm Mpc}^{-1}$) are extrapolated towards larger $k$
($\approx 1 \, h \: \ {\rm Mpc}^{-1}$).  As $\delta _k \propto k^{n/2}$,
the effect on $\sa$ is
\begin{equation}
 \frac{\sa (n)}{\sa (n=1)}\approx 10^{n-1}.
\end{equation} 
We see that a higher value of $n$ leads to a higher $\sa$.  We
emphasize that a value $n=1$ is not a prediction of all models of
inflation. In particular, a value of $n \approx 1.15$ has been
suggested in \cite{Wetterich:1989hf} for a natural explanation of the
smallness of density fluctuations by the long duration of inflation.
In this proposal, the density fluctuations on very small scales that
have left the horizon just at the end of inflation are of order unity.
The smallness of the inhomogeneities on galactic or even larger scales
is then explained by the slope in the spectrum and the `long lever
arm' once the corresponding scales left the horizon more than $50$ 
$e$-foldings before the end of inflation. A spectral index $n>1$
typically arises if inflation happens not too far from the
\emph{Planck} scale where effective couplings to higher order
curvature terms are still relevant. Such a scenario can typically be
found in dimensional models of inflation \cite{Shafi:1985ha}.

\section{The Maximum of the Power Spectrum}
As well as $\sa$, the presence of dark energy influences also other
features of the power spectrum, such as the location of its maximum or
the slope of decrease towards large $k$. In principle, appropriate
quantities related to these features could also be used to detect
quintessence.  At the moment it is not possible to locate the position
of the maximum of the power spectrum, which we denote here as $\kmax$.  
The spectrum is too
flat over a wide region, and the error bars are too large, however
observational data is improving \cite{twodf}.  The
$k$-value at which the power spectrum peaks, $\kmax$, is a very good
indicator for early quintessence but is rather insensitive to the
recent history of $w(a)$ (in contrast to $\sa$ which is sensitive to
both).  In the following, we will assume that $\omd$ is almost
constant during the early matter era, and we will identify the $\omd$
at that time with $\omdsf$.  We define here $\keq$ as the wave number
of the mode which enters the horizon just at matter-radiation
equality. From Equations \eqref{equality} and \eqref{age1} we get
\begin{equation}
 \keq=\frac{2 \pi}{\tau _{eq}}
 =0.165\: {\rm Mpc}^{-1}\times\left( \frac{h}{0.65}\right) ^2\frac{1 - \omd^0}
 {\sqrt{1-\bar{\Omega}_d (\aeq)}},
\end{equation}
where $\omdeq$ is given by  Equation \eqref{age2}.
We find that $\kmax$ is smaller than $\keq$  by a factor of more than four,
hence it enters the horizon during the early matter era. We define $\kappa$,
the ratio of $\keq$ to $\kmax$ 
\begin{equation}
 \kappa \equiv \frac{\keq}{\kmax}.
\end{equation}  
The slope of the power spectrum is roughly given by
\begin{equation}
 \frac{d \ln P(k)}{d \ln k}\approx n-4 \left ( 1-\frac{f_{\rm sub} (a_{\rm hor})}
 {f_{\rm sup}(a_{\rm hor})} \right ),
\end{equation}
where $f_{\rm sub}$ is the growth exponent on sub-horizon scales,
$f_{\rm sup}$ the one on super-horizon scales, and $a_{\rm hor}$ is
the scale factor when the mode $k$ enters the horizon. $f_{\rm sub}$
and $f_{\rm sup}$ depend only on the relative energy densities of the
$m$, $r$ and $d$ -component, but $\rho _r / \rho _m$ is always the
same at the same $a/\aeq$.  We conclude that $\kappa$ is only
sensitive to $\omdsf$ and the spectral index $n$.  We find that there
is also a slight dependence on the other parameters $h,\ \omdn$ and
$\Omega_{\rm b}$.  This is due to the baryons, which partially
suppress $f_{\rm sub}$ as long as they are coupled to the photons.
Hence the ratio $\aeq/\adec$ affects $\kappa$. One can easily find
fitting formulas for $\kappa$. For instance, models with $\omdn \in
[0.6, 0.7],\ h=0.65,\ n=1$ and $\Omega _b h^2=0.02$ are well described
by
\begin{equation}
 \ln \kappa = 1.57+2.70 \: \omdsf.
\end{equation}
Approximating $\bar{\Omega}_{\rm d}(\aeq) \approx \omdsf$ we find a strong dependence of 
the maximum of the power spectrum on early quintessence as for the above
parameters
\begin{equation}
\kmax \propto \left(1-\omdsf\right)^{-1/2} \exp(-2.7\, \omdsf) \approx (1-2.2\, \omdsf).
\end{equation}

\section{Specific Models}
As a typical model with early quintessence we choose the Leaping Kinetic Term (LKT)
model from \cite{Hebecker:2001zb}.\footnote{Another example of an early quintessence model
has been proposed in \cite{Albrecht:2000rm}}

\begin{figure}[!ht]
\begin{center}
\includegraphics[scale=0.4, angle =-90]{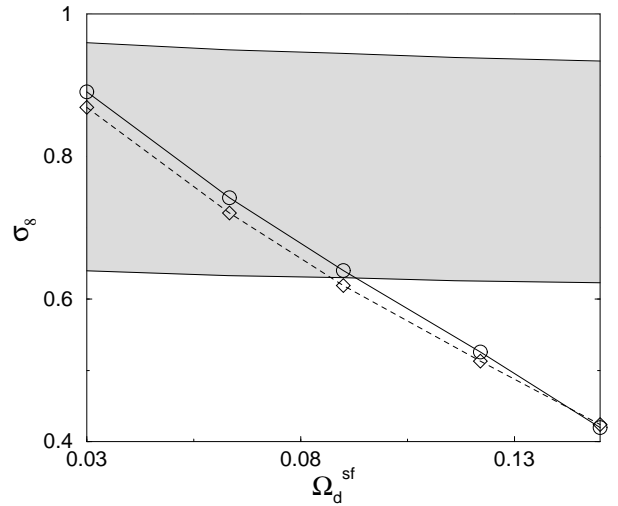} 
\caption{Numerical  value of $\sigma_8^{\rm cmb}$ (solid line) and analytic estimate from Equation \eqref{main} (dashed line)
for Leaping kinetic term quintessence. The cluster abundance constraint is shaded.
We use $h=0.7,\ \omdn=0.6,\ \Omega_{\rm b}h^2 =0.02,\ n=1$. }
\label{compare}
\end{center}
\end{figure}

\begin{figure*}[!ht]
\begin{center}
\subfigure[$h=0.6,\ \omd^0=0.6$]{
\includegraphics[scale=0.38, angle =-90]{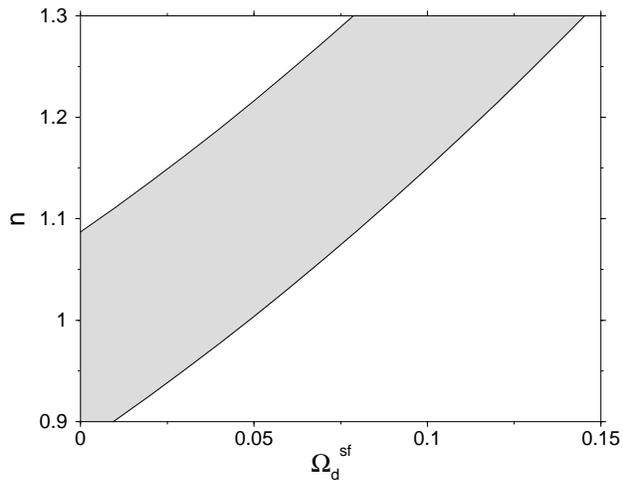} }
\subfigure[$h=0.6,\ \omd^0=0.7$]{
\includegraphics[scale=0.38, angle =-90]{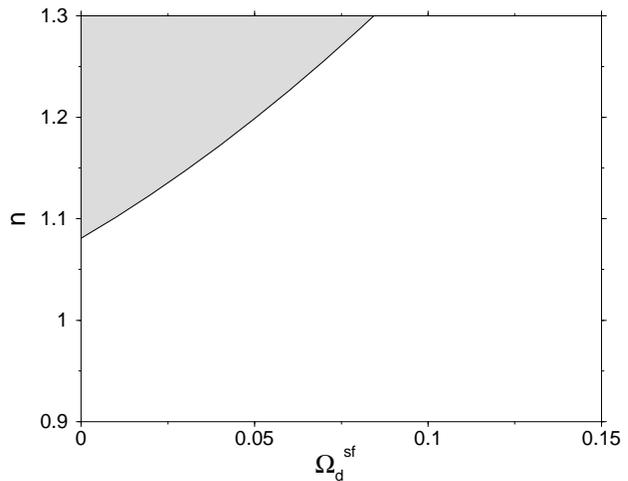} }
\subfigure[$h=0.7,\ \omd^0=0.6$]{ \label{forCmp}
\includegraphics[scale=0.38, angle =-90]{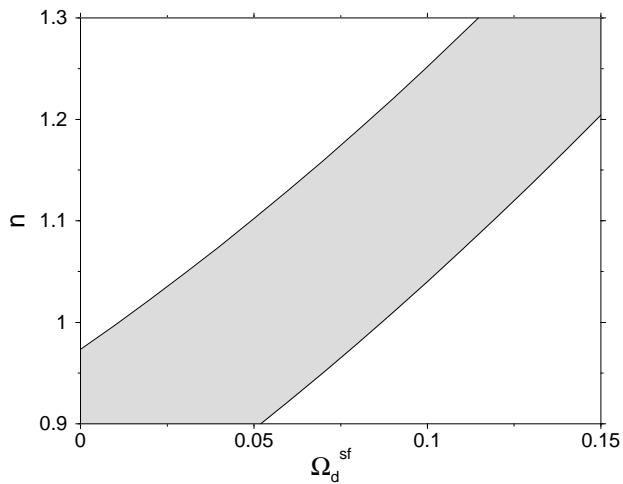} }
\subfigure[$h=0.7,\ \omd^0=0.7$]{
\includegraphics[scale=0.38, angle =-90]{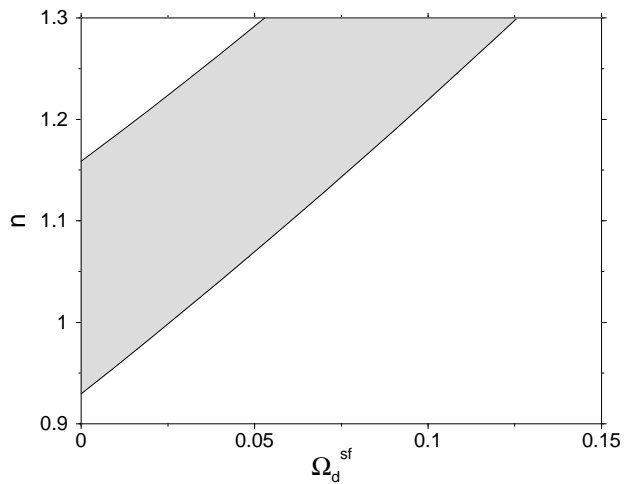} }
\end{center}
\caption{Allowed range of early quintessence and spectral index
$n$ for various values of the Hubble parameter $h$ and present
dark energy $\omd^0$.}
\label{range}
\end{figure*}

It has the Lagrangian
\begin{equation}
\mathcal{L}=\frac{1}{2}k^2(\phi)\partial _{\mu}\phi\partial^{\mu}\phi + e^{-\phi},
\end{equation}
\begin{equation}
 k(\phi)= 1+k_0 + \tanh (\phi-\phi _0),\ \phi _0 \approx 277,
\end{equation}
with $k_0$ directly related to $\omdsf$ by $\omdsf = 3 k_0^2$.
Figure \ref{compare} shows the $\omdsf$-dependence of the CMB normalized $\sa$ in this model
obtained by a numerical solution using a modified \mbox{\sc cmbfast} code. 
We find good agreement in comparison with 
 the analytic estimate Equation \eqref{main}, the agreement is excellent.
We observe a strong dependence of $\sa$ on the amount of dark energy during
structure formation, $\omdsf$. For the parameters used in 
figure \ref{compare} ($h =0.7,\ n=1,\ \Omega_{\rm b}h^2 =0.02,\ \omdn = 0.6$), a non-zero 
amount of early quintessence $\omdsf \approx  0.05 \pm 0.04$ would be favored (see also figure \ref{forCmp}).

We have also used the LKT model to explore the constraints on
quintessence \eqref{cluster}, being aware that there remain still some
theoretical and systematic uncertainties in estimate \eqref{cluster}.
We emphasize that according to our analytic discussion the use of specific
LKT models is not a restriction of generality. Other quintessence models with the same
values of $\omd^0,\ \omdsf$ and $\bar w$ will lead to the same
results.  In Figure \ref{range}, we have plotted the allowed range for
early time quintessence $\omdsf$ and the spectral index $n$ for
different choices of $\omd^0$ and $h$. A spectral index $n$ near unity
favors the absence of early quintessence if $\omd^0$ is large and $h$
is small.  On the contrary, for small $\omd^0$ and large $h$, a few
percent quintessence during structure formation are favored. This
holds, in particular if the spectral index is somewhat larger than
one, e.g.  $n\approx 1.15$ \cite{Wetterich:1989hf}.

An overall bound for early quintessence can be obtained from Equation \eqref{cluster}
if we assume $h< 0.75,\ \omd^0 > 0.5$ and $n < 1.2$. One finds
\begin{equation}
\omdsf \lessapprox 0.2
\end{equation}
This is of the same order as the bound from big bang nucleosynthesis,
$\Omega^{\rm bbn}_{\rm d} < 0.2$ \cite{Wetterich:1995bg,Birkel:1997py}. This bound 
can be substantially improved by more precise determinations of $h,\ \omd^0$ and $n$.

\section{Summary}
We have analyzed the effects of dark energy on structure formation. We found that
$\sa$ - and possibly $\kmax$ - are very promising indicators for constraining
the present amount of dark energy $\omdn$, the present equation of state $\wda$ and
especially the amount of early quintessence $\omdsf$. The CMB-normalized value of $\sa$ 
depends on all
cosmological parameters. As a rough guide for the strength of these dependencies 
around \emph{standard values} $\omdn = 0.65$, $h=0.65$, $n=1$,
$\Omega _b h^2=0.02$ with  $-1<\bar w<-0.5$ we get
\begin{itemize}
\item Increasing $h$ by 0.1 $\Rightarrow$ Increase of $\sigma _8$ by 20 \%
\item Increasing $\omdn$ by 0.1 $\Rightarrow$
      Decrease of $\sigma _8$ by 20\%   
\item Increasing $n$ by 0.1 $\Rightarrow$ Increase of $\sigma _8$ by 25\%
\item Increasing $\bar w$ by 0.1 $\Rightarrow$ Decrease of $\sa$ by 5-10\%
\item Increasing $\Omega _b h^2$ by 0.01 $\Rightarrow$
      Decrease of $\sigma _8$ by 10\%
\item Increasing $\omdsf$ by 0.1 $\Rightarrow$
      Decrease of $\sigma _8$ by 50\%
\end{itemize}
Comparing with observation, the dependencies listed can be used for a
quick check of viability for a given quintessence model and parameter
set.  
If $\omdn$ is increased by $0.1$, cluster abundances according to
Equation \ref{cluster} yield an approx. $20 \%$ higher value of $\sa^{\rm cluster}$.
In combination with the corresponding decrease of $\sa^{\rm cmb}$,
the net effect on the ratio $\sa^{\rm cmb} / \sa^{\rm cluster}$  
is therefore a decrease by $33 \%$.
For a $\Lambda$CDM universe with \emph{standard values} as above one has $\sa^{\rm cmb} =
0.90$ and $\sa^{\rm cmb} / \sa^{\rm cluster} = 1.01 \pm 0.2$.
Compatibility of the cosmological scenario requires this ratio to be
close to unity.

Once a subset of cosmological parameters such as $h,\ \omd^0,\ \Omega_{\rm b}h^2$ 
and $n$ are accurately determined by other measurements e.g.\ CMB 
anisotropies, the quantitative understanding of structure formation may become a central
ingredient for the distinction between various forms of dark energy. In
particular, by distinguishing quintessence from a cosmological constant
it could serve as an indicator for a new field, the cosmon, mediating a new
force with similar strength as gravity. If this field does not couple 
to ordinary matter, cosmology will be the only
way for proving or disproving its existence.\\[1ex]

\noindent{\bf \large Acknowledgments}\\[1ex]
We would like to thank Gert Aarts, Matthew Lilley and Eduard Thommes
for helpful discussions.

\end{document}